# The formation of strong electric fields and volumetric charges in the Solar atmosphere

**Sarsembaeva A.T.**

Physico-Technical Faculty, Theoretical and Nuclear Physics Department,
Al-Farabi Kazakh National University, Almaty 480012, Kazakhstan
aiganym@nucl.sci.hokudai.ac.jp

**Takibayev N.Zh.**

Al-Farabi Kazakh National University, Almaty 480012, Kazakhstan

**Kato. K.**

Division of Physics, Graduate School of Science, Hokkaido University,
Sapporo 060-0810, Japan

## Abstract

The processes occurring in the solar atmosphere are diverse and depend on many important factors. For example, from magnetic fields, their sudden changes, from emissions of substance from the depths of the Sun, distribution of shock waves and plasma jets, etc. The paper describes the model of formation of the charged volumes of gas in solar atmosphere, which is called solar "storm clouds" by analogy with terrestrial storm clouds. The model will be based on the theory ionization equilibrium and the Saha equation.

## I. Structure and properties of atmosphere of the Sun

Atmosphere of the planet - a gas shell of a celestial body, held near gravitation. Between the atmosphere and interplanetary space there is no sharp border. Atmosphere can be considered as the area around of the celestial body in which the gas environment rotates together with it as an entity. The depth of the atmosphere of some planets consisting in basically from the gases (gas planets), can be very large.

The solar atmosphere is heterogeneous and consists of several various layers (Fig.1). The deepest and thin of these layers is the photosphere. The thickness of the photosphere is about 300 kilometers which is observed in the visible continuous spectrum. Deeper layers of the photosphere have a higher temperature [4, 8, 9].

Consequence of the process of convection is the appearance of granules in the photosphere – sections whose dimensions are about a thousand kilometers; whose temperature is several hundred degrees higher than surrounding areas. Such "gas spots" create the impression that the photosphere has a cellular, "granular" structure.

The boundary area of the photosphere and chromosphere is the "cold" and is approximately $4000^0 K$. The temperature of the "top" and "down" from this area is growing rapidly.

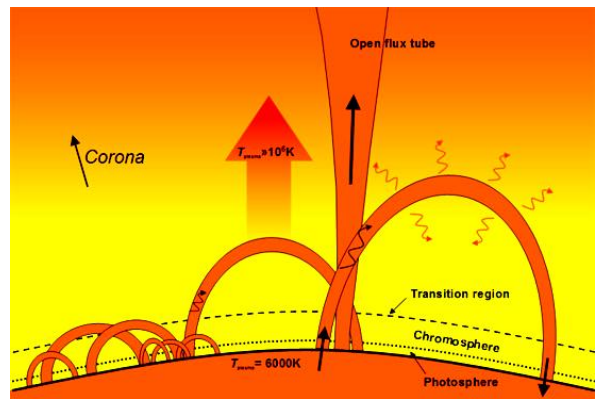

Fig. 1. The atmosphere of the Sun

Above the photosphere are more rarefied layers: the chromosphere and corona. They are almost transparent to the continuous optical radiation. Chromosphere in its structure is much more heterogeneous than the photosphere. We can distinguish two types of irregularities - bright and dark. Thus, the chromosphere has a slightly granular structure similarly to the photosphere, however heterogeneity of chromosphere larger than the granules of photosphere. Distribution of heterogeneity in the chromosphere forms the "chromospheric network". The temperature of the chromosphere in its upper part reaches about several tens of thousands of degrees.

Along the limb of the Sun brightness of the chromosphere is changing: in the active regions of an increasing number of the spicules - individual gas flows, and increased radiation. The average radiation of the chromosphere in active regions increases by 3-5 times, it corresponds to increasing the density of gas is approximately 2 times.

Chromosphere above 1500 km is essentially a set of relatively dense gas fibers, with density of particles $n_H \approx 10^{10}$-$10^{11}$ cm$^{-3}$ and temperature T $\approx$ 6000-15000 $^0$K, and jets with a much more rarefied gas between them. 4-5 thousand km above are only spicules [1].

In the cell of a chromospheric network gas spreads from the center to periphery at a speed of 0,3-0,4 km/sec. The magnetic field at the boundary cells is enhanced, average lifetime of such a formation - about a day. The horizontal spreading of the ionized gas from the center of a cell to periphery rakes a weak magnetic field (with vertical lines of force). Enhancement of the field is intensifying the glow of the chromosphere near the borders of the grid, just as it is in weak active regions [6].

The intensity of the radiation of the chromosphere as a whole is small. For stars of solar type found that the chromospheric emission lines in H, K, and etc. decreases with decreasing rotation speed of stars and their age. According to this criterion, the Sun – a star is quite old with low activity.

The most rarefied and hottest part of a solar atmosphere is the corona. Its temperature reaches a million degrees, the corona can be traced from the limb and to the distance of a few solar radiuses. Very important is the fact that the corona is visible only during a total solar eclipse; we can watch it by the special device – the coronagraph.

The whole solar atmosphere to be in constant oscillatory movements, both vertically and horizontally. These oscillations have a resonance character, and occur with a frequency of approximately once every 5 minutes. Because these oscillations is very strongly influenced by the magnetic fields of our luminary, in case of sharp increase in magnetic activity in some areas the Sun can be observed a surprising phenomenon – the corona prominences, flocculi in the chromosphere, facula and spots in photosphere. The most significant and large scale events (affecting all layers of the solar atmosphere) are solar flares.

## II. Formation of volumetric charges in atmosphere of the Sun

The processes occurring in the solar atmosphere are diverse and depend on many important factors. For example, from magnetic fields, their sudden changes, from emissions of substance from the depths of the Sun, distribution of shock waves and plasma jets, etc. Consider the model of formation of the charged volumes of gas in solar atmosphere, which is called solar "storm clouds". The model will be based on the theory ionization equilibrium and the Saha equation. They describe the equilibrium and the local quasi-equilibrium ionization of gases [5, 10, 15].

For example, consider what happens with counter flows of gas at the border of spicules in the chromosphere. Chromospheric spicules – individual gas streams, which rise or fall at a speed of about ~ 20 km/sec, differ by the fact that the gas streams are moving horizontally at a speed of more than 0,2 km/sec. from the center to the periphery of spicules. Thus, on the border of spicules is almost constant collision of gas flows. Furthermore, the colliding streams have different temperature – it follows from the data of the analysis of intensity and of the radiation spectrum of the chromospheric network [11].

In the counter flows of gas - plasma substance at the boundary of the chromospheric cells in the border area between streams occur a collision of atoms and molecules of the oncoming flows, that is resulting for increased dissociation molecules and ionization of atoms and molecules.

In this area – an area of "friction" of gas streams, generated positive and negative ions, mainly it is hydrogen radicals $H_2^+, H_2^-, H^-$, and some of the free electrons [3].

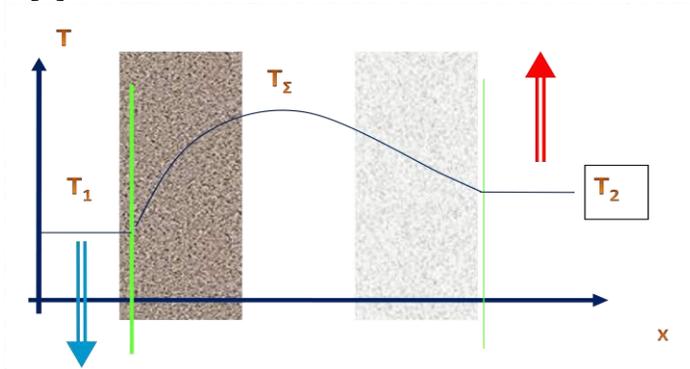

Fig. 2. Temperature characteristics of area of the collision of two counter flows with different temperatures $T_1$ and $T_2$. The x axis is perpendicular to the flow. Double arrows indicate on counter motion of flows. In the area of "friction" darker square denotes cold area, and by lighter – a hot area.

If the counter flows have a different temperature the friction zone acquires a temperature gradient from the center of the zone, where the temperature is maximal to its borders (Fig. 2). This is the typical pattern associated with the movement and interaction of two streams of cold and warm layers of an atmosphere. Heterogeneity and turbulent distortions in the area of "friction" for simplicity we omitted

For a variety temperature gradients from the center of "friction" in a direction to area 1 and a direction to area of 2 concentration of ions of different types will be different. Indeed, the concentration of ions of each grade has its well-defined temperature dependence. This is a consequence of the balance of direct and reverse reactions which intensity depends on temperature. This balance is described by Saha which often called the theory of ionization equilibrium [5].

Note that the from data of ionization energies of atoms and molecules of hydrogen and the Saha equation, it follows, that in the colder flows will be concentrated mainly positive ion molecule $H_2^+$, and in warmer masses - negatively charged ions $H^-$.

## III. Method ionization equilibrium

Consider a general thermodynamic description of the reactions occurring in gaseous medium. General conditions of chemical equilibrium of reactions follows from the law of mass action

$$\sum \nu_i \mu_i = 0 \tag{1}$$

Where $\mu_i$ – chemical potentials of the reacting substances, and $\nu_i$ – balance factors. For example, for the reaction of formation water molecules

$$2H_2 + O_2 - 2H_2O = 0$$

balance factors are equal $\nu(H_2) = 2$, $\nu(O_2) = 1$, $\nu(H_2O) = -2$. This reaction has explosive character in normal conditions of nature – the explosion of a hydrogen-oxygen gas mixture.

In general the reactive component may be several. The chemical potential of each gas mixture is equal to [10]

$$\mu_i = T \ln P_i + \chi_i(T), \tag{2}$$

Where $P_i$ – partial pressure of the gas of number i, $P_i = C_i \cdot P$, P - total pressure of the gas mixture, $C_i = N_i / N$ - the concentration associated with the ratio of the number of particles of a grade $N_i$ to the total number of particles of the gas mixture: 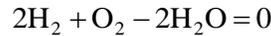 $N = \sum N_i$.

Chemical equilibrium constant is given by

$$K_p(T) = \exp(-1/T \sum \nu_i \mu_i). \tag{3}$$

Here and below, chosen system of units: $c = 1, \hbar = 1$ and the Boltzmann constant $k = 1$. This expression is directly related to the well-known formula of the law of mass action

$$\prod C_i^{\nu_i} = P^{-\sum \nu_i} K_P(T) \equiv K_C(P,T).$$

(4)

Note, the law of mass action is carried out for the reactions between the dissolved substances, if their concentrations in the solvent are small [3].

The law also valid for the reactions of ionization and recombination of atoms. This situation also takes place for processes in weakly ionized gases.

## IV. Saha equation

The problem of the «ionization equilibrium» monoatomic plasma is as follows. We assume, that at low temperatures the gas consists of neutral atoms. Then, with increasing temperature the atoms ionized by collisions and recombination radiations: single - $A_0 \to A_1 + e^-$ double- $A_1 \to A_2 + e^-$, triple- $A_2 \to A_3 + e^-$, etc. Here as usually, have introduced the notation: for neutral atom - $A_0$, singly ionized atom - $A_1$, doubly ionized - $A_2$, etc. [2].

Actually, the ionization equilibrium is a special case of chemical equilibrium. In application to these reactions the law of mass action leads to system of the equations

$$\frac{C_{n-1}}{C_n C} = P K_p^{(n)}(T) , \qquad n = 1, 2, \ldots \qquad (5)$$

where $C_0$ - concentration of neutral atoms, $C_1$, $C_2$,... - concentration of ions with different degrees, and $C$ - concentration of free electrons.

Equality $C = C_1 + 2 \cdot C_2 + 3 \cdot C_3 + \cdots$ reflects the electrical neutrality ionized gas as a whole. The equilibrium constants for monatomic gases are easily identified. Heat capacity of the particles are equal $c_P = 5/2$, and then

$$K_P^n = \frac{g_{n-1}}{2g_n} \frac{1}{T^{5/2}} \left(\frac{2\pi}{m_e}\right)^{3/2} \exp(I_n/T) , \qquad (6)$$

where $g = (2L+1)(2S+1)$ - statistical weight of atoms or ions (L, S – their orbital angular momentum and spin), $m_e$ - electron mass, $I_n = \varepsilon_{0,n} - \varepsilon_{0,n-1}$ - energy of the n-th ionization.

Reduced system of equations (5) determines the concentration of various ions and gives their functional dependence with change of temperature (the formula of Saha) [3].

When increasing the temperature, increasing the number of ionized atoms, and when the temperature of ions and electrons recombine and form neutral atoms. Concentrations are usually defined as the ratio of the number of particles in this grade to the total number of particles in the medium [16].
## V. The ionization of weakly ionized hydrogen plasma

Consider the ionization of gas molecules in the solar atmosphere at different temperatures.
In collisions of hydrogen molecules occur as dissociation and ionization. We write, for example, reactions that require a significant amount of energy on

$$H + Q \to H^+ + e \qquad Q \geq 13{,}6\ eV\ ;$$
$$H_2 + Q \to H_2^+ + e \qquad Q \geq 15{,}4\ eV\ ;$$
(7)

where Q - energy derived by a molecule in collisions with particles of the medium or γ - rays.

We write the reactions involving the hydrogen molecule, requiring less energy

$$H_2 + Q \to H + H \qquad Q \geq 4{,}5\ eV\ ;$$
$$H_2 + H_2 + Q \to H_2^+ + H_2^- \qquad Q \geq 3.922\ eV\ ;$$
(8)

and

$$H_2^+ + Q \to H + H^+ \qquad Q \geq 2.652\,eV\ ;$$
$$H_2 + H + Q \to H_2^+ + H^- \qquad Q \geq 1.062\,eV\ ;$$

(9)

as well as pickup reaction with a second electron (the reaction is endothermic)

$$H + e + Q \to H^- \qquad Q \leq -0.754\,eV\ . \qquad (10)$$

The ionization processes, considered as the chemical equilibrium reaction, the temperature dependences of ion concentrations are defined by similar equilibrium constants - $K^{(n)}(T)$. It should be noted, of course, additional internal degrees of freedom in the diatomic hydrogen molecules and molecular ions [14].
For the negative $N_{H^-}(T)$ ion concentration of the $H^-$ hydrogen atom should

$$\frac{N_{H^-}}{N_H \cdot N_e} = \frac{g_{H^-}}{g_H \cdot g_e} \left(\frac{2\pi}{m_e T}\right)^{3/2} \exp\left(\frac{\varepsilon(H^-)}{T}\right), \qquad (11)$$

where $\varepsilon(H^-)$ - binding energy of the electron in the state $(1s^2\,1S_0)$ of an ion $H^-$, $g_{H^-} = 1$ - the statistical weight of this state, $g_H = 2$ - the statistical weight of the hydrogen atom in the state $(1s)$, $g_e = 2$ - the statistical weight of the electron [3].

For concentration $N_{H_2^+}(T)$ of the positive molecular ion $H_2^+$ formula of Saha will have the form

$$\frac{N_{H_2^+}}{N_H \cdot N_{H^+}} = \frac{Z(H_2^+)}{g_H \cdot g_{H^+}} \left(\frac{2\pi}{m_e T}\right)^{3/2} \exp\left(\frac{D}{T}\right), \qquad (12)$$

where $D = 2{,}65\,eV$ – the energy of dissociation $H_2^+ \to H + H^+$, $g_{H^+} = 1$, and $Z(H_2^+)$ - is the internal statistical sum of the molecular ion $H_2^+$, which takes into account the sum of all the rotational and oscillatory levels $H_2^+(X\ ^2\Sigma_g^+)$ [7, 12, 13].

Layers at the border itself flows and areas of "friction" are areas of partial mutual penetration of charged particles, i.e. ions, ionic molecules and electrons.

## VI.  Model of formation of "storm clouds" in the solar atmosphere

Consider the flow in chromosphere of the Sun. Analysis of observations and calculations have shown, that even at a temperature of 6000 °K concentration ratio of the negative hydrogen ion and a neutral hydrogen atom is equal $N_{H^-}/N_H \approx 10^{-8}$ [12], i.e. weakly ionized gas.

At temperatures T ~ 5000 - 15000 °K the total concentration of molecular hydrogen ions $H_2^+$ is of the same order as the total concentration of ions $H^-$. However, with decreasing temperature, in particular, at T < 3000-3500 °K, the concentration $N_{H_2^+}$ is much greater than $N_{H^-}$ (Fig.3). It should be noted that even at high temperatures the concentration of neutral hydrogen molecules $H_2$ in the ground electronic state to several orders magnitude higher than the concentration of ions $H_2^+$. For example, in the photosphere

$$N_{H_2}/N_{H_2^+} \approx 10^4 - 10^5 \ .$$

Data on the temperature dependences of the relative concentration of various ions of hydrogen are given in Sobelman work with co-authors [12].

For us important that even in low ionization temperature differences in the concentrations of ions of different charges, in particular, ions $H_2^+$ and $H^-$, lead to the enrichment of ions $H_2^+$ in layers in the area of the friction with the lower temperature, and ions $H^-$ of layers of this area, having a heat temperature.

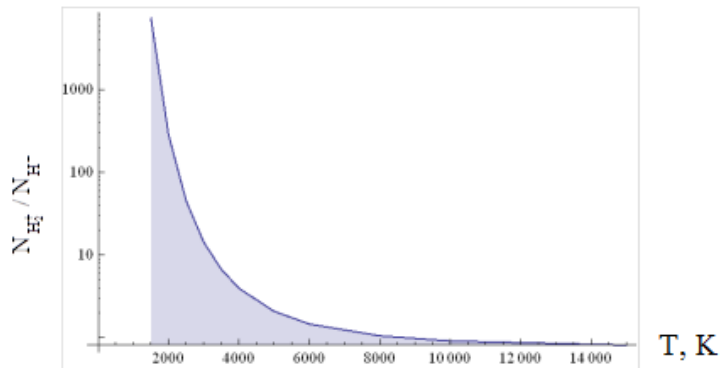

Fig. 3. The ratio of relative concentration of $N_{H_2^+}/N_{H^-}$ dependence on the temperature of equilibrium hydrogen plasma.

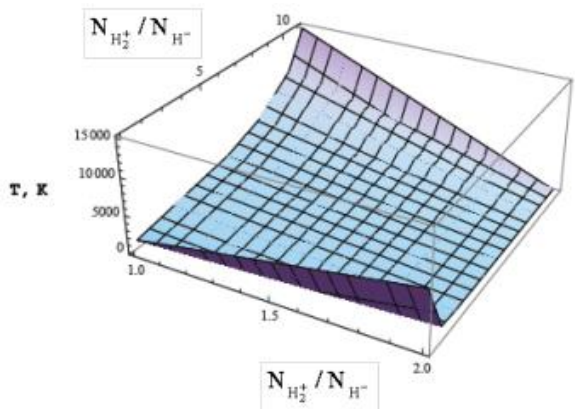

Fig. 4. The 3D graphic of the Fig.3.

Fig. 5 shows the flow and local areas with different temperatures. A flow going from left to right, has, for example, the temperature $T_1$ much lower than the temperature $T_2$, which the flow going from right to left.

In the ionization zone and friction can be turbulent motions and plasma oscillations of free electrons relative to positive ions. Some of these electrons will be trapped by ions or atoms in that part of the zone where conditions for their capture would be advantageous.

Thus, in the scheme of Figure 5, the bottom of the friction zone and ionization temperature is higher and more effectively producing H- which will penetrate and accumulate in a flow with temperature $T_2$.

At the top of the friction zone temperature is below and more efficiently produced molecular hydrogen ions $H_2^+$ which will penetrate and accumulate in the stream at lower temperature $T_1$.

We will discuss the physical cause of the solar electrification of «storm clouds». In the central area of friction the temperature and a degree of ionization will be the highest. It will consist of gas molecules and hydrogen atoms and ions: $H^-$, $H_2^+$, $H_2^-$, as well as free electrons, as a whole is electrically neutral. Free electrons is still very small, and they will be quickly trapped by atoms and positive ions [3].

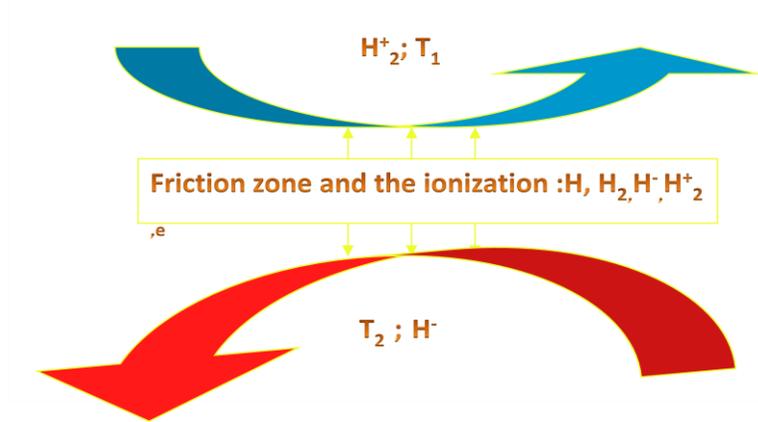

Fig. 5 The scheme of motion of colliding streams with different temperature. Shows the direction of increasing concentrations of ions having different temperature dependences (enrichment of layers of oppositely charged ions in the field of "friction").

In the layers closer to the region 2 with a temperature $T_2 > T_1$ of hydrogen atoms $H^-$ ions will be carried away by the flow more than the friction of heavy molecules.

In the layers closer to the region 1 is more fractions of the heavy particles – hydrogen molecules and molecular ions $H_2^+$ and $H_2^-$. Note in this connection that, according to energy balance of reactions of (16) and (12) advantage in the concentration ionic molecules $H_2^+$ will be obvious before the ionic molecule $H_2^-$.

Thus, the mass of opposite flow - diffusion, thermodiffusion and thermodynamic effects lead to a gradual separation of the charged fractions, and creating singular gas-plasma «electric dynamo».

During the movement counter flows will be enriched by ions of opposite charges, and moving away from each other, will create a large volumetric of uncompensated value of the charges.

Formed "charged clouds" will form in the solar atmosphere a complex system of electric and magnetic fields, supplementing the electric and magnetic fields from the other structures of the Sun and its atmosphere.

Using data from the temperature dependences of ion concentrations can be estimated the total uncompensated volumetric charge of the Sun "storm clouds". They are formed in the colliding gas flows on the borders in the chromospheres spiciles of the Sun. We consider that these uncompensated charges in divergent flows are equal, but opposite in sign.

The volumetric uncompensated charge in the clouds is the product of a unit charge to the total number of ions (two and multiple ionization will be considered small in magnitude)

$$Q = e \cdot N_{H^-}, \qquad (13)$$

where e - unit charge, $N_{H^-}$ the total sum of ions in the gas stream. This sum can be determined, knowing the sizes and flow temperature, and ion concentration of the given grade.

From the observations and spectral analysis follows, that

$$b_H = \frac{N_{H^-}}{N_H} = 10^{-8}$$

at temperature $T = 6000^0 K$. The total number of neutral atoms in a flow $N_H = n \cdot V_O$, where n - number of atoms per unit of volume (concentration of atoms), $V_O$ - volume of cloud (flow) of hydrogen gas. On the heights of the chromosphere, we assume for example, that the concentration of gas particles is $n \approx 2{,}7 \cdot 10^{17}$ m$^{-3}$, and the size of the gas cloud is

$$V_O = 10^4 \cdot 500 \cdot 500 \, \text{km}^3 = 2{,}5 \cdot 10^{18} \, \text{m}^3,$$

i.e. a small cloud by solar scopes.

When the temperature difference of flows

$$\Delta T = T_2 - T_1 \approx 3000 \ ^0 K,$$

estimates give a huge number of negative ions in the «storm cloud» (Fig. 5)

$$N_{H^-} \approx 10^{-8} \cdot N_H \approx 6{,}7 \cdot 10^{35} \qquad (14)$$

and huge on size uncompensated volumetric charges, carried by each of the streams

$$Q = e \cdot N_{H^-} \approx 10^{17} \, C. \qquad (15)$$

These «storm clouds» will form strong network of electric fields in the solar atmosphere, and with the motion of flows - will form strong magnetic fields. An original cause of the energy of these fields is giant energy of rotation of the Sun and thermal energy which emanating from its depths [3].

For accurate calculations, of course, should take into account several important factors. This is turbulence of flows, the influence of magnetic fields, the influence of fluctuation of the density associated with shock and periodic waves in the solar atmosphere, effect of the radiation flux, etc.

Unevenness of the eruption of matter and radiation from the Sun, changing the complex structure of magnetic fields and convective flows create conditions for the permanent oppositely formation for the charged gas streams. Their categories - «solar lightning», is a permanently ongoing processes which serve as background to well-known flares and active regions in the solar atmosphere.

## VII. Conclusion

This article was reviewed by some of the diverse and interesting phenomena occurring in the solar atmosphere. We have touched upon only those them that have been associated with the formation and development of volumetric atmospheric electrical charges and their features.

On the Sun we call them «storm clouds» by analogy with terrestrial storm clouds. They are carriers of large magnitude of electric charges, which occurring lightnings - electrical discharges.

In the solar atmosphere also occur lightnings, i.e. electrical discharges, emissions of plasma (spicules, coronal mass ejections and flares, etc.).

Have been made estimations the magnitude of the electric charge of the vast masses of gas flows in the solar atmosphere, having different temperatures and, often, opposite of the direction of motion. It was clear, that the huge energy of motion of these masses is transformed by collisions to the huge potential of the electric field, i.e. uncompensated electric charges collected in the three-dimensional structure – gas plasma clouds in the Sun.

The important things is that the concentrations of different ions have different temperature dependence. The reason for this is their electronic structure, i.e. especially the internal quantum mechanic structure of the energy levels of these ions and ionic molecules. But important to the development of macroscopic atmospheric processes is not only microscopic characteristics of ions (such as energy of ionization or dissociation, their mass and molecular structure, the moments, oscillatory and rotational spectra, etc.), but also macroscopic characteristics of gas flows, external fields and radiation.

As the temperature changes the concentration of some ions can significantly exceed the concentration of other ions, even if the concentration at a given temperature were equal to each other, it gives a key to understanding of why volumetric charges can be separated in space.

There are important features of motion flows relative to each other, which have different temperatures. First, if flows do not have relative motion, the difference of temperatures between them cannot long persist. Second, if relative motion is small, the large volumetric charges of opposite sign cannot be deducted any extended time away from each other. And only, if the relative motion of flows is sufficient in magnitude, then it can spread the volumetric charges away from each other and maintain their own temperature in each of the streams almost unchanged.

Locally isolated volumetric and electrically charged object – a "the storm cloud" can be discharged, when it met with another cloud, which having an opposite charge and emerged in a similar way elsewhere. An example of this, in our opinion, can serve as a terrestrial storm clouds and the behavior of gas flows in the spicules in the chromospheres of the Sun.

## Acknowledgements

This work was supported financially in part by Al-Farabi Kazakh National University. S.A. and T.N. would like to thank the University of Hokkaido for hospitality and support during theirs stay in Sapporo.

During my study was signed Academic Exchange Agreement between Al-Farabi Kazakh National University and Hokkaido University.

We also thank all members of Theoretical Nuclear Physics Laboratory of Hokkaido University for their support during the preparation of this paper, for helpful discussions and comments.